\documentstyle[preprint,prd,aps]{revtex}
\tightenlines
\begin{document} 
\title{\Large A stationary relativistic representation of the bound state}  
\author{L.Micu\thanks{E-mail address: 
lmicu@theory.nipne.ro}} 
\address{Department of
Theoretical Physics\\ Horia Hulubei Institute for Physics and Nuclear
Engineering\\
 Bucharest POB MG-6, 76900 Romania}
\maketitle
\begin{abstract}
We show that a bound system in momentum space can be treated like a
gas of free elementary constituents and a collective excitation of a
background field which represents the countless quantum fluctuations generating
the binding potential. The distribution function of the internal momenta in
the bound system at rest is given by the projection  of a solution of a
relativistic bound state equation on the free wave functions of the elementary 
constituents. The 4-momentum carried by the collective excitation is the
difference between the bound state 4-momentum and the sum of the free
4-momenta. This definition ensures the explicit fulfilment of Lorentz
covariance, mass-shell constraints and single particle normalizability of the
bound state function.  The discussion
is made for a two particle bound state and can be easily generalized to the
case of three or more particles.
\end{abstract}

This paper is an attempt to supply the acute need of a relativistic treatment
of the bound state problem manifested in particle physics.

In the following we shall specifically refer to the meson as a bound state
of a quark and an antiquark, although the method is quite general. For reasons
of simplicity we shall omit the flavour and colour indices and assume that the
interaction potential is white.

In the standard approach of the bound state problem as it is found in
quantum mechanics, the existence of a bound state is conditioned by the
presence of an attractive potential well. The wave function is stationary and
normalizable in the space of the relative coordinates with respect to the
center of forces. We remind however that the interaction potential is
essentially a nonrelativistic notion and hence this approach bears a
nonrelativistic character even if one uses the relativistic expression of the
energy in the time independent dynamical equation \cite{gi}.

In the relativistic approaches derived from field theory the binding is
supposed a consequence of quantum fluctuations, that is of a continuous
exchange of quanta among the constituents \cite{bs}. The iterative solution
can be expressed in terms of free propagators and an interaction kernel which
makes the number of elementary constituents indefinite. In spite of some formal
similarities with the nonrelativistic quantum mechanical treatment of the
bound state problem this approach is of a very different type. This is
reflected in the presence of a relative time coordinate, in the ambiguities
of the definition of the relative momentum, in the perturbative definition of
the interaction kernel, in the existence of negative norm states. For these
reasons the function describing the inner structure of a bound state is
similar to a form factor rather than to a wave function. This situation is
significantly improved by the elimination of the relative time coordinate in
quasipotential models which resort to a relativistic
extension of the Schr\"odinger equation for the bound state where the
interaction potential is replaced by a scattering kernel \cite{qp}. 

For the understanding of the relation between the quantum
mechanical and the field approach it has been essential to know if a bound
system with a fixed number of particles can be quantized in a relativistic
manner. Dirac \cite{pamd} has shown that this is indeed possible if the
generators of the symmetry group depend explicitly on the interaction
potential. He also has shown that by restraining the symmetry group to those
transformations which are purely kinematical it is possible to develop
dynamical models independent on the concrete form of the interaction. This
idea stays at the origin of the light cone models \cite{lc}. The full
relativistic covariance is of course destroyed in this case, but the effects
of symmetry violation may be evaluated by comparing various quantization
schemes \cite{as}.

The purpose of the present paper is to create a link between the quantum
mechanical and the quantum field approaches of the bound state problem 
this time starting from a relativistic equation for a bound
state. Our specific aim is to get a Lorentz covariant representation of
the  interaction potential compatible with the field approach, which will
help in obtaining a real relativistic representation of a meson as bound
state. The work is done in the momentum representation which is adequate to our
purpose and the constituents of the bound state are treated as independent
particles. 

We assume that the total Hamiltonian is the sum of two
free Dirac Hamiltonians and of an interaction potential which depends on both
coordinates. Its eigenfunctions are the internal functions of the bound system
in the rest frame and the corresponding eigenvalues are the meson masses. We
then have: 

\begin{equation}\label{h}
[\sum_{j=1,2}( -i\vec{\nabla}^{(j)}\vec{\alpha}^{(j)}+\beta m_j)+ {\mathcal
V}_0(\vec{x}_1,\vec{x}_2)]\Psi(\vec{x}_1,\vec{x}_2)=M
~\Psi(\vec{x}_1,\vec{x}_2) 
\end{equation}
where $\alpha$ and $\beta$ are Dirac matrices and ${\mathcal V}_0$ is the
interaction potential. The function $\Psi(\vec{x}_1,\vec{x}_2)$ has two
spinorial indices and can be best written under the form of a $4\times4$
matrix. 
 
For reasons related to the real independence of the quarks which will become
clear below we assume that the wave function $ \Psi(\vec{x}_1,\vec{x}_2)$
satisfies the set of constraints 

\begin{equation}\label{p}
 (-i\vec{\nabla}^{(1)}-i\vec{\nabla}^{(2)}+\vec{\mathcal
V}(\vec{x}_1,\vec{x}_2)) \Psi(\vec{x}_1,\vec{x}_2)=0
\end{equation}
where $\vec{\mathcal V}(\vec{x}_1,\vec{x}_2)$ is a vector operator. Its
expression as well as the compatibility
between eq.(\ref{h}) and the eqs.(\ref{p}) will be briefly commented in the
next in agreement with the interpretation we give to the interaction
potential. (N.B. If $\vec{\mathcal V}=0,~\Psi(\vec{x}_1,\vec{x}_2)\sim
\Psi(\vec{x}_1-\vec{x}_2)$ and the real independence of the quarks is
lost.) 

In the space representation the solution of eqs.(\ref{h}) and (\ref{p})
corresponding to the eigenvalue $M_{\{n\}}$ where $\{n\}$ is a set of quantum
numbers labelling the bound state is denoted by 
$\Psi_{\{n\}}(\vec{x}_1,\vec{x}_2)=\left\langle \vec{x}_1,\vec{x}_2\vert
\Psi_{\{n\}}\right\rangle$. Its projection on the solutions of the free
Dirac equation $\psi_{\{\vec{k}_j\}}(\vec{x}_j)={\rm exp}(i\epsilon_j\vec{k}_j
\vec{x}_j) w(\{\vec{k}_j\})$ and $\psi^c=
C\bar\psi^T_{\{\vec{k}_j\}}(\vec{x}_j)$ is denoted by $\bar{w}_i(\{\vec{k}_1\})
\left(\Psi_{\{n\}}(\{\vec{k}_1\},\{\vec{k}_2\})\right)_{ij}
\bar{w^c}^T_j(\{\vec{k}_2\})$, where $C$ is the charge conjugation
matrix, $w$ is a Dirac spinor and $\{\vec{k}_j\}$ represents the set of quantum
numbers $\{\epsilon_j,~ s_j,~\vec{k}_j\}$ labelling the free states
($\epsilon_j=\pm$ is the sign of the energy, $s_j$ is the projection of the
spin on an arbitrary axis and $\vec{k}_j$ is the momentum). 

It is an easy matter to see that the projection satisfies the following set
of equations derived from (\ref{h}) and (\ref{p}) 

\begin{eqnarray}\label{eqs1}
\left(\epsilon_1\sqrt{\vec{k}_1^2+m_1^2}+\epsilon_2\sqrt{\vec{k}_2^2+m_2^2}
\right)
&\bar{w}&(\{\vec{k}_1\})\Psi_{\{n\}}(\{\vec{k}_1\},\{\vec{k}_2\})
\bar{w^c}^T(\{\vec{k}_2\})+\langle \{\vec{k}_1\},\{\vec{k}_2\} \vert{\mathcal
V}_0\vert\{n\}\rangle\nonumber\\
&=&
M_{\{n\}}~\bar{w}(\{\vec{k}_1\})
\Psi_{\{n\}}(\{\vec{k}_1\},\{\vec{k}_2\})\bar{w^c}^T(\{\vec{k}_2\})
\end{eqnarray}
\begin{equation}\label{eqs2}
(\epsilon_1\vec{k}_1+\epsilon_2\vec{k}_2)
\bar{w}(\{\vec{k}_1\})\Psi_{\{n\}}(\{\vec{k}_1\},\{\vec{k}_2\})
w(\{\vec{k}_2\})+ \langle \{\vec{k}_1\},\{\vec{k}_2\}\vert\vec{\mathcal
V}\vert\{n\}\rangle=0. 
\end{equation}

According to the general principles of quantum
mechanics the projection $\bar{w}(\{\vec{k}_1\})
\Psi_{\{n\}}(\{\vec{k}_1\},\{\vec{k}_2\})\bar{w^c}^T(\{\vec{k}_2\})$ is the
probability amplitude for finding two free quarks with the individual quantum
numbers $\{\vec{k}_1\}$ and $\{\vec{k}_2\}$ in the meson state characterized
by $\Psi_{\{n\}}$.
Eqs.(\ref{eqs1}) and (\ref{eqs2}) show however that the representation of
the meson as a supperposition of two free quark states is incomplete,
because the sum of the quark 4-momenta does not satisfy the meson mass-shell
constraint. A real representation must include the contribution of the
interaction potential in such a way as to preserve relativistic
covariance.    

The solution we found to this problem was to assume the existence of a
third component of the meson beside the valence quarks and independent of
them. This component is denoted by $\Phi$ and carries the 4-momentum $Q^\mu$
\cite{micu} which is the difference between the meson and the quark momenta:

\begin{equation}\label{Q}
Q^0=M_{\{n\}}-\epsilon_1\sqrt{\vec{k}_1^2+m_1^2}-
\epsilon_2\sqrt{\vec{k}_2^2+m_2^2} 
\end{equation}
\begin{equation}\label{vQ}
\vec{Q}=-\epsilon_1\vec{k}_1-\epsilon_2\vec{k}_2.
\end{equation}

We notice that $\Phi(Q)$ does not have a definite mass and hence it is not an
elementary excitation of the quark gluonic field. 

We are now allowed to represent
the meson   by a gas of free quarks having the distribution of momenta and
spins given by  $\bar{w}(\{\vec{k}_1\})\Psi_{\{n\}}(\vec{k}_1,\vec{k}_2)
\bar{w^c}^T(\{\vec{k}_2\}) $ {\it and} a collective excitation of the
background field, carrying the missing momentum $Q^\mu$ defined in (\ref{Q})
and (\ref{vQ}). Then one can write the single meson state at rest in the
following form:

\begin{eqnarray}\label{meson}
&&\left.\vert {\cal M}_{\{n\}}(M_{\{n\}},0)\right\rangle=
\int d^3k_1~{m_1\over e_1}
d^3k_2{m_2\over e_2}d^4Q 
~\delta^{(3)}(\vec{k}_1+\vec{k}_2+\vec{Q})\delta(e_1+e_2+Q_0-M_{\{n\}})
\nonumber\\
&&\times \sum_{s_1,s_2}
\bar{u}_{s_1}(\vec{k}_1)\Psi_{\{n\}}(\vec{k}_1,\vec{k}_2)C v_{s_2}(\vec{k}_2)
~\Phi^\dagger(Q)~  a^\dagger_{s_1}
(\vec{k}_1)b^\dagger_{s_2} (\vec{k}_2)\vert 0\rangle 
\end{eqnarray}
where $\{n\}$ are the quantum numbers of the meson, $a^\dagger$ and
$b^\dagger$ are quark and
antiquark creation operators and $u,v$ are free Dirac
spinors. 
We mention that only the projection of $\Psi_{\{n\}}$
on the positive energy states appear in the expression of the single meson
state (\ref{meson}) because those corresponding to the negative energy are
associated with the quark and antiquark annihilation operators which gives 0
when acting on the vacuum. This eliminates the complications due to the
presence of negative energy states which in alternative
approaches of the bound state problem have been cured by the introduction of
positive energy projectors in the definition of the bound state Hamiltonian
\cite{hs}. 
We remark however that the situation changes when two or more mesons are
present because not all the annihilation operators are now acting on the
vacuum. They may annihilate the quarks in other mesons giving rise to a
genuine form of interaction among the bound systems usually called "pair
creation-annihilation mechanism".

It is worthwhile noticing here that eqs.(\ref{h}),
(\ref{p}) and (\ref{meson}) can be immediately generalized to the baryon case
by introducing a third quark contribution according to
the general rules of relativistic covariance. 

The physical significance of $\Phi$ can be deduced from the
correspondence between the quantum mechanical and the field approaches.
Specifically, observing that $Q^\mu$ represents the contribution
of the interaction potential to the bound state momentum we conclude that
$\Phi^\dagger$ is the collective,
time averaged effect of the continuous series of virtual excitations of the
quark gluonic field giving rise to the binding. $Q^0$ is then a kind of binding
energy and $\vec{Q}$ is the reaction of the whole mass of virtual particles to
the motion of the free valence quarks or, in other words, it is the effect of
the imperfect cancellation of the vector momenta during the quantum
fluctuations. In order to clarify some concrete aspects of the relation among
$\vec{Q}$ and $\vec{\mathcal V}$ we consider the case where
the quarks are  independently bound to the center of forces, like for instance
in the bag model \cite{bm}. We assume accordingly that the quark momenta are
uncorrelated and hence eqs.(\ref{vQ}) and (\ref{p}) are identities defining
the bag momentum and the operator associated to it respectively. In this case
one can write $\vec{\mathcal V}(\vec{x}_1,\vec{x}_2)=
i\vec{\nabla}^{(1)}+i\vec{\nabla}^{(2)}$ solving in this way the
problem of compatibility  among eqs.(\ref{h}) and (\ref{p}). Obviously this
is not the unique choice for $\vec{\mathcal V}$ but it is the simplest
compatible with our assumptions.

Closing this discussion on the significance of the collective excitation
$\Phi$ we notice that it has some common features with the bag \cite{bm}. Both
$\Phi$ and the bag are effective, nonelementary extra components of
the bound system representing the binding effects. We also remark the
resemblance between the definition of the momentum $Q^\mu$ (see
eqs.(\ref{Q}),(\ref{vQ})) and that of the effective potential in
nonrelativistic QCD, as the  remaining part in the effective Lagrangian
after removing off the kinetic terms \cite{bb,pot}. 

A last comment on the expression (\ref{meson}) concerns the form of
$\Psi_{\{n\}}(\{\vec{k}_1\},\{\vec{k}_2\})$ which is a 4$\times$4 matrix
describing the distribution of the internal momenta and the coupling of the
quark spins and angular momenta in the meson. $\Psi_{\{n\}}$ can then be
written as a linear combination of Dirac matrices with coefficients behaving at
rotations like the components of a tensor. From general arguments related to
the transformation properties of the meson wave function one can write for
instance: 

\begin{eqnarray}
&&(\Psi_{P}C)_{lm}=
\varphi_P\gamma^5_{lm}\nonumber\\
&&(\Psi_{V}C)_{lm}=\varepsilon^i\left(
\varphi^V_{1}
\gamma^i_{lm}+\sum_{n=1,2}\varphi^V_{2n}k_n^i\delta_{lm}
+\sum_{n=1,2}\varphi^V_{3n}[\gamma^i,\gamma^j]_{lm}k_n^j
\right)\nonumber\\
&&(\Psi_AC)_{lm}=
\varepsilon^i\left(\varphi^A_{1}\gamma^i_{lp}+
\sum_{n=1,2}\varphi^A_{2n}k_n^i\delta_{lp}
+\sum_{n=1,2}\varphi^A_{3n}[\gamma^i,\gamma^j]_{lp}k^j_n\right)\gamma^5_{pm}
\end{eqnarray}
where $P,~V,~A$ denote the pseudoscalar, vector and axial mesons
respectively, $\varepsilon$ are the meson polarization vectors having only
spatial components in the rest frame and $\varphi_i$ are
scalar functions of $\vec{k}_1,\vec{k}_2$ whose arguments have been omitted for
simplicity. 

We are now ready to put the expression (\ref{meson}) in a Lorentz covariant
form simply by replacing  $\vec{k}$ by $k^\mu_T$ and the scalar product
$\vec{k}_i\vec{k}_j$ by  $-k_{iT}^\mu k_{jT\mu}$ where
$k_T=(0,\vec{k})$ in the rest frame. Under the change of reference
frame $k_T$ transforms into 
$k^{'\mu}_T=\Lambda(\vec{\omega})^\mu_\nu k^\nu_T$ and the Dirac spinors
$u,v$ transform into $u(\vec{k}')=\Lambda_{sp}(\vec{\omega})u(\vec{k})$ where
$\Lambda(\vec{\omega})$ and $\Lambda_{sp}(\vec{\omega})$ are the
vectorial and spinorial representations of Lorentz transformations from the
rest frame to a reference frame moving with the velocity $\vec{\omega}$ with
respect to the first one. The result is the expression of the meson state with
the energy $E={M\over\sqrt{1-\vec{\omega}^2}}$ and the momentum
$\vec{P}=-\vec{\omega}~E$:

\begin{eqnarray}\label{m1}
&&\left.\vert {\cal M}_{\{n\}}(E,\vec{P})\right\rangle=
\int d^3k_1~{m_1\over e_1}
d^3k_2{m_2\over e_2}d^4Q 
~\delta^{(3)}(\vec{k}_1+\vec{k}_2+\vec{Q}-\vec{P})\delta(e_1+e_2+Q_0-E)
\nonumber\\
&&\times \sum_{s_1,s_2}
\bar{u}_{s_1}(\vec{k}_1)\Psi_{\{n\}}(k_{1T},k_{2T})C
v_{s_2}(\vec{k}_2) ~\Phi^\dagger(Q)~  a^\dagger_{s_1}
(\vec{k}_1)b^\dagger_{s_2} (\vec{k}_2)\vert 0\rangle 
\end{eqnarray}
It must be emphasized that the expression (\ref{m1}) is free of the
ambiguities in the definition of the quark momenta appearing in the models
derived from quantum field theory. This fact and the explicit fulfilment of
the mass shell constraints both by the meson and by the quark
4-momenta would be impossible in the absence of the nonelementary
excitation $\Phi$. This one carries the "missing
4-momentum" (see eqs. (\ref{Q}),(\ref{vQ})) which is just what one needs to be
added to the free quark momenta in order to get the meson momentum. 

To complete the proof we have still to verify the
normalizability of the single meson state (\ref{meson}). To this end we make
use of the commutation relations of the free quark operators and define the 
vacuum expectation value of the collective excitation in such a way as to 
ensure the separate conservation of its 4-momentum:

\begin{equation}\label{vev}
\left\langle 0\right\vert~\Phi(Q)~\Phi^+(Q')~\left\vert 0 
\right\rangle~={1\over V_0T_0}~
\int d^4~X~{\rm e}^{i~(Q'-Q)_\mu~X^\mu}=
{(2\pi)^4\over L^3_0T_0}~\delta^{(4)}(Q-Q')
\end{equation}
where $1/V_0~T_0$  has been introduced for dimensional reasons with
$V_0$ the meson volume and $T_0$ a time sensibly larger than the time
basis involved in the definition of the collective excitation $\Phi$.

In order to avoid the
cumbersome $\delta(E-E')$ at the expression of the norm induced by
$\delta(Q_0-Q'_0)$ in (\ref{vev})  and to preserve in the same time the 
manifest Lorentz covariance we write

\begin{eqnarray}\label{delta} 
{2\pi\over T_0}\delta(Q_0-Q'_0)
&=&{1\over T_0}\int_T dX_0{\rm e}^{i(Q_0-Q'_0)X_0}
={1\over T_0}\int_T dX_0{\rm e}^{i(E(P)-E(P'))X_0}
\approx \nonumber\\
&&{E\over M_{\{n\}}T_0}\int_{T_0}
dY_0 {\rm e}^{i(M_{\{n\}}-M_{\{n'\}})Y_0}\approx{E\over
M_{\{n\}}}\delta_{M_{\{n\}}M_{\{n'\}}} 
\end{eqnarray}
and get immediately:

\begin{equation}\label{norm}
\left\langle~{\cal M}_{\{n'\}}(P')~\vert {\cal
M}_{\{n\}}(P)~\right\rangle~=
2E~(2\pi)^3~\delta^{(3)}(P-P')~\delta_{M_{\{n\}}M_{\{n'\}}}
~\delta_{\{n\}\{n'\}}~{\cal J} 
\end{equation}
where

\begin{eqnarray}\label{J2} 
&&{\cal J}=~{1\over 2M~L_0^3}~\int d^3k_1~{m_1\over
e_1}~d^3k_2~{~m_2\over e_2}~d^4Q~
\delta^{(4)}(k_1+k_2+Q-P)\nonumber\\
&&\times Tr\left({\hat k_1+m_1\over 2m_1}~\Psi_{\{n\}}(k_{1T},k_{2T})C
{-\hat
k_2+m_2\over2m_2}~\gamma^0 C^+\Psi^+_{\{n'\}}(k_{1T},k_{2T})\gamma^0\right)=
\delta_{\{n\}\{n'\}}. \end{eqnarray}

In the above relations we have implicitly assumed that $M_{\{n\}}$ and
$M_{\{n'\}}$ are discrete eigenvalues of the equation (\ref{h}) with 
$\vert M_{\{n\}}-M_{\{n'\}}\vert~T_0>>1$ so that the integral in (\ref{delta})
vanishes if $M_{\{n\}}\ne M_{\{n'\}}$. We notice that relation (\ref{delta})
allows to eliminate the rather
arbitrary time $T_0$ from the expression of the norm, which is quite 
remarkable. 

The normalization relation (\ref{norm}) can also be seen as
an expression of the confinement because it shows
that a many particle state (\ref{meson}) is
normalized like a single particle one if the integral $\cal J$
converges. This will not be possible if $Q^\mu=\alpha P^\mu$ because
${\mathcal J}$ would contain the highly singular factor
$\delta(\alpha-\alpha')\delta^{(3)}(0)$. This is a decisive
argument for introducing the constraints (\ref{p}) with $\vec{\mathcal V}\ne0$
which guarantees the existence of a nonvanishing momentum $\vec{Q}$ in the
rest frame of the meson. 

The last point we discuss here is the way back from the field
representation (\ref{meson}) to the wave function $
\Psi(\vec{x}_1,\vec{x}_2)$ in order to see if it is possible to recover
the  last one from the first.

The calculations are performed in the meson rest
frame, where the bound state wave function has been defined. We notice 
that only the degrees of freedom associated with the quarks have a meaning in 
quantum mechanics. Those associated with the collective excitation do not
and must be integrated out. Then, by analogy with quantum field theory and
having in mind the stationarity of the meson structure, we define its wave
function as follows:      

\begin{equation}\label{wf}
\tilde\Psi_{\{n\}}(\vec{x}_1,\vec{x}_2,t)_{\alpha \beta}= 
\langle 0\vert\int d^3Q~\bar\psi^c(\vec{x}_2,t)_{\beta}
\bar\psi(\vec{x}_1,t)_{\alpha}\Phi(\vec{Q},t)\vert{\cal
M}_{\{n\}}(M_{\{n\}},0)\rangle  
\end{equation} 
where the single meson state is given by (\ref{meson}), $\psi$ is the free
quark field, $\alpha$ and $\beta$ are spinorial indices and  
\begin{equation}
\Phi(\vec{Q},t)={T_0\over(2\pi)}\int~dQ_0 {\mathrm
e}^{-iQ_0t} \Phi(\vec{Q},Q_0).
\end{equation}

By straightforward calculations it can be seen that the time
dependence factorizes out under the form  ${\rm e}^{-iM_{\{n\}}t}$ (see
eq.(\ref{Q})) and that
$\tilde\Psi_{\{n\}}(\vec{x}_1,\vec{x}_2,t)_{\alpha \beta}$ contains only that
part of the initial function  $\Psi_{\{n\}}(\vec{x}_1,\vec{x}_2)$ having
nonvanishing projection on the positive energy free states. The part
projecting on the  negative energy free states is lost, because it does not
appear in the definition of the meson state (\ref{meson}). 

In conclusion, the expression (\ref{meson}) may be seen as a link between the
quantum mechanical and a field representation of a bound state because it
describes in field language the information obtained in relativistic quantum
mechanics. This has been achieved by assuming the existence of a nonelementary
effective component, besides the valence quarks, which in quantum mechanics can
be related to the interaction potential and in quantum field theory can be
seen as a collective excitation of some background field. The conclusion is
that a real relativistic representation of a bound state is possible in
momentum representation where stationarity, Lorentz covariance and mass shell
constraints can be simultaneously and explicitly satisfied. 

\vskip0.5cm     
{\bf Acknowledgments}
This work was started during author's visit at ITP of
the University of Bern in the frame
of the Institutional Partnership Program of the Swiss National
Science  Foundation under Contract No.7 IP 051219. The
author thanks Prof. Heiri Leutwyler for hospitality and stimulating 
discussions.  

A careful reading of the manuscript and suggestions by Fl.
Stancu as well as clarifying discussions with Irinel
Caprini are also gratefully acknowledged.


\begin{thebibliography}{99} 

\bibitem{gi}
D. P. Stanley and D. Robson, Phys. Rev. D {\bf 21}, 3180 (1980);
S. Godfrey and N. Isgur, Phys. Rev. D {\bf 32}, 189 (1985);
N. Isgur, D. Scora, B. Grinstein and M. B. Wise, Phys. Rev. D {\bf
39}, 799 (1989); N. Barik, B. K. Dash and M. Das, Phys. Rev. D {\bf
31}, 1652 (1985); N. Barik, P. C. Dash, A. R. Panda, Phys. Rev. D
{\bf 46}, 3856 (1992).

\bibitem{bs} 
E. E. Salpeter and H. A. Bethe, Phys. Rev. {\bf 84}, 1232 (1951);
M. Gell-Mann and F. Low, Phys. Rev. {\bf 84}, 350 (1951); N.
Nakanishi, Suppl. Prog. Theor. Phys. {\bf 43}, 1 (1969).

\bibitem{qp}
A. A. Logunov and  A. N. Tavkhelidze, Nuovo Cim. {\bf 29}, 380 (1963);
V. G. Kadyshevsky, Nucl. Phys. {\bf B6}, 125 (1968); F. Gross, Phys.
Rev. {\bf 186}, 1448 (1969); F. Gross and J. Milana, Phys. Rev. D {\bf
43}, 2401 (1991); R. N. Faustov, V. O. Galkin and A. Yu. Mishurov,
Phys. Rev. D {\bf  53}, 6302 (1996); R. N. Faustov and V. O. Galkin,
Phys. Rev. D {\bf 52}, 5131 (1995). 

\bibitem{pamd}
P. A. M. Dirac, Rev. Mod. Phys. {\bf 21}, 392 (1949).

\bibitem{lc}
H. Leutwyler and J. Stern, Ann. Phys. (N. Y.) {\bf 112}, 94 (1978);
G. P. Lepage and S. J. Brodsky, Phys. Rev. {\bf D22}, 2157 (1980);
P. L. Chung, F. Coester and W. N. Polizou, Phys. Lett. {\bf B 205},
545 (1988); 
A. Szczepaniak, Chueng-Ryong Ji and S. R. Cotanch, Phys. Rev. {\bf D49}, 3466
(1994); S. J. Brodsky, hep-ph/0004211,

\bibitem{as}
A. Szczepaniak, A. G. Williams, Phys. Rev. {\bf D47}, 1175 (1993).

\bibitem {micu}
D. Ghilencea and L. Micu, 
Phys. Rev. D {\bf 52}, 1577 (1995);
L. Micu, Phys. Rev. D {\bf 53}, 5318 (1996); {\it ibidem} D {\bf 55},
4151 (1997). 

\bibitem{hs}
G. Hardekopf and J. Sucher, Phys. Rev. A {\bf 30}, 703 (1984).

\bibitem {bm}
A. Chodos, R. L. Jaffe, K. Johnson, C. B. Thorn and V. F. Weisskopf,
Phys. Rev. D {\bf 9}, 3471 (1974);
A. Chodos, R. L. Jaffe, K. Johnson and C. B. Thorn, Phys. Rev. D
{\bf 10}, 2599 (1974); T. DeGrand, R. L. Jaffe, K. Johnson and J.
Kiskis, Phys. Rev. D {\bf 12}, 2060 (1975).

\bibitem{bb}
N. Brambilla, hep-ph/0008279; G. S. Bali, hep-ph/0010032;
N. Brambilla and A. Vairo, Phys. Rev. D {\bf 55}, 3974 (1997); 
N. Brambilla, P. Consoli and G. M. Prosperi, Phys. Rev. D {\bf 50}, 5878  
(1994). 

\bibitem{pot}
K. G. Wilson, Phys. Rev. D {\bf 10}, 2445 (1974);
L. S. Brown and W. I. Weisberger, Phys. Rev. D {\bf 20}, 3239 (1979).

\end{thebibliography}
\end{document}